# Dynamic Defense Approach for Adversarial Robustness in Deep Neural Networks via Stochastic Ensemble Smoothed Model

Ruoxi Qin[1#], Linyuan Wang[1#], Xingyuan Chen[2], Xuehui Du[1], Bin Yan[1*]

[1]Henan Key Laboratory of Imaging and Intelligent Processing, PLA Strategy Support Force Information Engineering University, Zhengzhou, China

[2] PLA Strategy Support Force Information Engineering University, Zhengzhou, China

**Abstract**: Deep neural networks have been shown to suffer from critical vulnerabilities under adversarial attacks. This phenomenon stimulated the creation of different attack and defense strategies similar to those adopted in cyberspace security. The dependence of such strategies on attack and defense mechanisms makes the associated algorithms on both sides appear as closely reciprocating processes. The defense strategies are particularly passive in these processes, and enhancing initiative of such strategies can be an effective way to get out of this arms race. Inspired by the dynamic defense approach in cyberspace, this paper builds upon stochastic ensemble smoothing based on defense method of random smoothing and model ensemble. Proposed method employs network architecture and smoothing parameters as ensemble attributes, and dynamically change attribute-based ensemble model before every inference prediction request. The proposed method handles the extreme transferability and vulnerability of ensemble models under white-box attacks. Experimental comparison of *ASR-vs-distortion* curves with different attack scenarios shows that even the attacker with the highest attack capability cannot easily exceed the attack success rate associated with the ensemble smoothed model, especially under untargeted attacks.

## 1. Introduction

Deep learning approaches have been successfully employed in various computer vision applications, ranging from image detection [1] and image classification [2] to face recognition [3] and autonomous driving [4]. With such a wide range of applications, deep learning is particularly suited and accustomed to contemporary every-day life. In 2014, Szegedy et al. [5] found that the learned input-to-output mappings of deep deep neural networks are largely discontinuous, to the extent that small perturbations in some network inputs can lead to high misclassification errors. Such network inputs are considered to be adversarial samples, whose emergence raises concerns about the CNN applicability and reliability. Also, numerous studies addressed the CNN robustness and resilience to attacks in several applications. Subsequently, many adversarial learning methods, which are similar to cyberspace security ones, have been devised for both the attack and defense sides.

Methods of attack primarily focus on adversarial samples because of their proactive role in attack and defense scenarios. Attack algorithms can be divided based on their capabilities into white-box and black-box methods [5]. On the one hand, white-box methods are mainly based on a complete knowledge of the network gradients. For example, the fast gradient sign method (FGSM) [6] is a basic and effective method in which adversarial samples are generated by adding the sign reverse gradient to the original images in a constrained fashion. Also, the basic iterative method (BIM) [7] extends the single-step FGSM by using small-step iterations. As well, the momentum iterative method (MIM) [8] adds a momentum term in each iteration step to stabilize the updated direction and improve the transferability of the adversarial samples. Furthermore, the projected gradient descent method (PGD) [9] uses a first-order distortion term based on a BIM random start to improve the transferability. In addition to improving transferability, distortion reduction is also a key consideration in the design of attack algorithms. For example, the DeepFool [10] method uses the sample-to-hyperplane distance to achieve minimum-distortion search. Also, the C&W attack [11] uses a Lagrangian formulation for distortion reduction. On the other hand, black-box attacks have no knowledge of the network gradients. These attack methods can be divided into score-based and decision-based methods according to the attacker's model observations. The Natural Evolution Strategies (NES) [12] and Simultaneous Perturbation Stochastic Approximation (SPSA) [13] methods are score-based methods that employ random sample gradient estimation, while NATTACK [14] generates adversarial samples from a Gaussian distribution centered at the input sample. The boundary attack [15] is a decision-based method that uses a random search step on the decision boundary.

[#] First co-author  
[*] Corresponding Author

The rapid development of attack methods is also constantly interspersed with the proposal of defense methods. The two types of methods act as two sides of a competitive game developed in a mutually-promoting reciprocating process. Research on the defense methods started from a theoretical investigation of the existence of adversarial samples. The network vulnerability caused by the linear nature of high-dimensional image features [6], [16] has been captured and exploited by many attack methods. On the contrary, many empirical and certified defense methods seek to reduce this vulnerability and obtain robust networks. The most flexible and effective empirical defense method is the adversarial training associated with the generation of adversarial samples [9], [17]. The robust network performance through such training depends on the adversarial sample generation method. Indeed, every defense mechanism is effective and flexible for a specific attack method. Although empirical defense methods are convenient, their applicability is practically limited, because the associated adversarial robustness depends on the attack method and hence the attacker could generate more challenging adversarial samples [13], [18]–[20]. For instance, the distillation method [21] can employ gradient shielding to successfully foil traditional white-box attacks, but this method becomes ineffective under the high transferability of the C&W attack [11]. The model ensemble [21, 22] was firstly proposed as a defense method but has been proved ineffective [22] and used instead as an attack method that improves the transferability of adversarial samples [23]. The nature and wide applicability of empirical defense methods have stimulated intense increasing competition with attack methods, where the defense methods are mostly passive.

Certified defense methods are supported by rigorous theoretical security guarantees, where a robustness radius is obtained under the $l_p$ distortion constraint. These methods overcome the shortcomings of empirical defense methods through exact or conservative approaches. The exact methods are usually based on the satisfiability modulo theory [24] or linear programming [25] but they show limited scalability for large datasets [26]. The conservative methods show relatively better computational efficiency at the expense of fluctuating performance [27]. The certified defense methods can largely break down the attacks with rigorous theoretical guarantees. Nevertheless, these certified defense methods are still not widely used with CNN architectures on big data. Recently, theoretical guarantees for CNN security has been gradually combined with relevant aspects in cyberspace security. The random smoothing method, based on differential privacy [28], was originally used as an empirical defense method, and its correctness was gradually proved by a limited theoretical guarantee [29], [30].

Both empirical and certified defense approaches have definite limitations. Following the theoretical developments of cyberspace security, the two sides in a competitive game with no strongly secure defense method will eventually reach a Nash equilibrium [31]. Accordingly, generalized robust-control defense methods, namely the moving target defense (MTD) [32] and the dynamic defense model (DDM) [33], [34], were proposed with probabilistic formulations of the network attributes. The nature of randomness and unpredictability of the model makes it more difficult for the attacker to detect so that defense the attacks, which makes up for the existing defences theory and approach. This paper adopts the DDM ideology in deep neural network, and proposes an attribute-based stochastic ensemble model. Through this method, the attribute of the ensemble model change dynamically before every inference prediction request so that make the model difficult to be detected and predicted for attacker. The attributes of the proposed model include the model order, the network architecture, and the smoothing parameters. These attributes are randomly adjusted based on a heterogeneous redundancy model collection to dynamically structure ensemble model with different gradient at each time inference prediction request. Hence, the adjusted stochastic ensemble model with probabilistic attributes shows more favorable characteristics of diversity, randomness, and dynamics. To a greater extent, the fixed correspondence between input and output on the gradient is changed. The unpredictability of gradient makes it impossible for attackers to directly implement white-box attacks and the inefficient universality of the adversarial samples make the attacker helpless. Figure 1 shows a conceptual diagram of this dynamic defense method proposed in this paper.

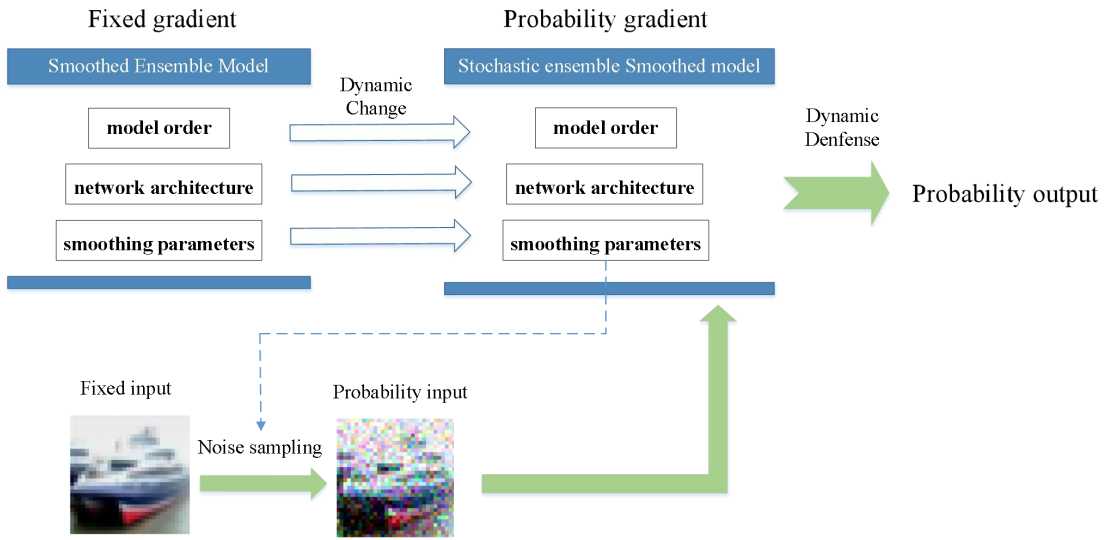

FIGURE 1. A graphical illustration of the proposed dynamic defense strategy for adversarial robustness of deep neural networks.

## 2. Related work

Random smoothing originally proposed based on the intention of differential privacy so that prevent the attackers from getting the exact gradient information for adversarial example [28].

This gradient shielding method based on input transformation is also a dynamic defense method intuitively. Random smoothing has been combined with adversarial training [35] to improve the certified defense flexibility. Indeed, limited dynamic defense outcomes are achieved by the combination with random smoothing which exploits the random noise perturbation of a network input. However, those method only provides dynamics on the input, and because it is based on a specific noise distribution, the subsequent attack method can still successfully attack under white-box attack through the expectation of input.

The complementary advantages of both empirical and certified defense methods can be obtained through combining these methods into new hybrid ones. Random smoothing has also been combined with the model ensemble [36] to enhance the provable robustness radius. From a dynamic perspective, the improvement of this defense performance is because the dynamics of input are amplified by the random input of multi candidate models, but ensemble model exists as a fixed architecture, and the dynamics are still limited to the input of the model.

Limited dynamics of model still face the threat of white box attacks under the expected estimation. However, these aforementioned methods have not essentially ended the afore-mentioned arms race. Apart from earlier combined defense methods, our proposed method creates a more heterogeneous and redundant model collection, and dynamically account for variations in ensemble attribute such as model quantity, network architecture, and smoothing parameters. The ensemble attribute will dynamically change and structure different ensemble model at each time inference prediction request to achieve the dynamic change of ensemble model. In summary, our work has the following main contributions:

(1) The deep network attributes are treated as variable attributes in the ensemble model, a heterogeneous redundant smoothed network collection is established, and the best attributes are stochastically selected in order to achieve superior dynamic defense performance.
(2) The random nature of model prediction is considered, the potential defense risk under attack is assessed, and the adversarial robustness is evaluated through Monte Carlo simulations and curves of attack success rate versus distortion (*ASR-vs-distortion*).
(3) Attackers with different capabilities are carefully designed under different dynamic defense scenarios, in order to comprehensively and intuitively evaluate the adversarial robustness of the proposed model in such scenarios.

## 3. Materials and Methods

### 3.1 Preliminaries of stochastic ensemble modeling

Let the random smoothing model *g* be trained by a basic classifier *f* through sampling, adding the noise $\delta \sim N(0,\sigma^2 I)$ to the input images, and minimizing the classification loss. The output of smoothed model *g* is defined as a mathematical expectation as follows:

$$g(x) = \mathrm{E}_{\delta \sim N(0,\sigma^2 I)}\left[f(x+\delta)\right]. \tag{1}$$

The model ensemble obtains the final prediction as the sum of the function outputs of the candidate models. The ensemble model $f_{ens}$ containing K numbers of models is defined as

follows:

$$f_{ens}(x,\theta) = \sum_{k=1}^{K} f(x,\theta_k).  \qquad (2)$$

The Smoothed WEighted ENsemble (SWEEN) approach creates an ensemble smoothed model with a weight parameter $\omega$ for each model, and improves the provable robustness radius [36]. In terms of the probability distribution of the input noise, the predicted output of the SWEEN model is given by a mathematical expectation operator as follows:

$$SWEEN = \mathrm{E}_\delta\left[\sum_{k=1}^{K}\omega_k f(x+\delta;\theta_k)\right] = \sum_{k=1}^{K}\omega_k \mathrm{E}_\delta\left[f(x+\delta;\theta_k)\right] = \sum_{k=1}^{K}\omega_k g(x;\theta_k). \qquad (3)$$

The constant weight parameters $\omega$ of the candidate models are independent of the SWEEN model output and can be optimized as $\omega^*$. Different from SWEEN, the ensemble attributes of proposed stochastic ensemble model (SEM) are randomly adjusted to dynamically structure ensemble model at each time inference prediction request making the output of candidate models in SEM has an additional mathematical expectation in terms of probability occurrence. However, the probability of occurrence of a certain candidate model under the SEM is assumed to be determined by the expectation $\mathrm{E}(f_k)_{occurrence} = \omega_k$ .and statistically independent from prediction expectation. Hence, the stochastic ensemble and SWEEN models can be equivalent in terms of output expectations. The theoretical improvement of the robustness radius by the SWEEN model is a special case of that of the SEM, which improves the dynamic properties of the ensemble, and achieves a more generalized dynamic change of model gradient in each inference prediction:

$$\begin{aligned}\mathrm{SEM} &= \mathrm{E}\left[\sum_{k=1}^{K} f_k(x+\delta;\theta_k)\right] = \sum_{k=1}^{K}\mathrm{E}(f_k)_{apparence} \times E\left[f_k(x+\delta;\theta_k)\right]\\ \mathrm{SEM} &= \sum_{k=1}^{K}\omega_k^* E\left[f_k(x+\delta;\theta_k)\right] = \sum_{k=1}^{K}\omega_k E\left[f_k(x+\delta;\theta_k)\right] = \mathrm{SWEEN} \quad \text{when } \omega_k = \omega_k^*\end{aligned} \qquad (4)$$

3.2 Heterogeneous redundant model collection based on random smoothing

Inspired by the moving target defense (MTD) method, randomness, diversity and dynamics are essential for robust model control. So, a heterogeneous redundant model collection is introduced as an essential and novel part of our dynamic defense method. The heterogeneity of the model architectures provides more randomized gradient information for the ensemble, while the redundancy of the classification models in the collection leads to model architecture diversity that maintains good classification performance. For the stochastic ensemble method, the model quantity, the network architecture, and the smoothing parameters are employed as variable ensemble attributes.

The stochastic ensemble method is intuitively used for gradient shielding under white-box attacks. The gradient uncertainty of ensemble model based on dynamically change of ensemble attribute before each time inference prediction request leads to a different gradient direction for each generated adversarial sample, and hence results in confusion about the update direction of that sample. In order to ensure a large range of gradient directions and meet the diverse needs of the stochastic ensemble, each individual model in the ensemble needs to have widely varied gradient information. Other studies also pointed out the key role of this diversity in ensemble defense methods [37]. Random smoothing is a robust training method in which the network input is perturbed with random noise. Because random smoothing is independent of the architecture and has a diversified selection of noise parameters, this method is effective in expanding the diversity and redundancy of the model collection.

The heterogeneous redundant model collection in this paper includes architectures of different depths and widths, where random smoothing is conducted on all models in the collection through a pre-trained model [47]. As shown in Table 1, a Monte Carlo simulation is used to compute the *approximated certified accuracy* (ACA) for the prediction performance. Although some simple models (like AlexNet and shallow VGG) cannot achieve stable smoothed prediction, the unsmoothed models are used for stochastic ensemble smoothing. The experimental results shall demonstrate the key role of the heterogeneity of the model collection on the robustness of the stochastic ensemble models.

TABLE I Results of the heterogeneous redundant model collection and the smoothed models on CIFAR10

| Model architecture | Smoothing parameter $\sigma$ | | | Model architecture | Smoothing parameter $\sigma$ | | |
|---|---|---|---|---|---|---|---|
| | 0.25 | 0.75 | 1.5 | | 0.25 | 0.75 | 1.5 |
| **DenseNet** [38] | | | | **VGG** [39] | | | |
| DenseNet100（95.5） | 94.03 | 89.96 | 83.56 | VGG11（92.1） | 9.99 | 80.11 | 20.88 |
| DenseNet121（94.1） | 91.23 | 87.01 | 82.08 | VGG13（94.3） | 65.67 | 10.0 | 61.18 |
| DenseNet161（94.2） | 92.31 | 87.88 | 82.80 | VGG16（93.9） | 9.99 | 9.99 | 9.99 |
| DenseNet169（94.0） | 91.29 | 87.96 | 81.11 | VGG19（93.3） | 91.83 | 87.50 | 81.74 |
| **WRN** [40]（96.2） | 91.78 | 90.23 | 83.43 | **AlexNet** [41]（77.2） | 9.99 | 9.99 | 9.99 |
| **ResNet** [42] | | | | **InceptionV3** [43]（93.8） | 91.91 | 86.86 | 80.38 |
| ResNet18（93.3） | 90.49 | 86.63 | 80.15 | **MobileNetV2**[44]（94.2） | 88.91 | 84.74 | 77.35 |
| ResNet34（92.9） | 91.20 | 87.20 | 81.76 | **ResNext** [45]（96.2） | 93.12 | 88.70 | 80.62 |
| ResNet50（93.9） | 91.16 | 86.29 | 80.28 | **GoogleNet** [46]（92.7） | 91.63 | 87.61 | 80.64 |

3.3 Stochastic ensemble smoothing with variable attributes

In a model ensemble, the dynamic attribute changes and gradient variations in each smoothed model lead to temporal gradient variations. This paper proposes a dynamic defense method based on a heterogeneous redundant model collection and a stochastic ensemble strategy.

The SEM randomness is reflected in the randomness of the model attributes. Specifically, when the SEM is queried multiple times for gradient or output information, the ensemble is achieved through a completely stochastic choice of the model quantity, network architecture, and smoothing parameters in each query. Each ensemble-based prediction shall first perform a stochastic selection of the ensemble attributes. This selection process is divided into four steps: (a) Determine the model quantity for the stochastic ensemble; (b) Randomly select the number of model architectures among the seven available ones according to previous step; (c) Randomly select different smoothing parameters $\sigma$ for each model architecture; (d) Perform ensemble and inference prediction for each candidate model. Figure 2 shows a flowchart of the stochastic ensemble strategy.

By including the model architecture into the ensemble attributes, we allowed each iteration of the ensemble training to have gradient differences based on the network architecture changes. Furthermore, the network depth and smoothing parameters were exploited as ensemble attributes to increase the ensemble diversity in the gradient direction. The model quantity in each ensemble iteration is indeed small compared to all those in the model collection library to create gradient differentiation. On the one hand, a larger number of model quantities in each ensemble iteration will reduce the ensemble diversity and gradient variations. On the other hand, large ensemble quantity will lead to enhanced transferability of adversarial samples generated under a single ensemble iteration. For a probabilistic ensemble, using a single model will not affect the mathematical expectation of the prediction and can ensure a gradient change in each ensemble iteration. The model quantity attribute has a key role and an important influence on the method robustness as will be discussed in detail in the experimental analysis.

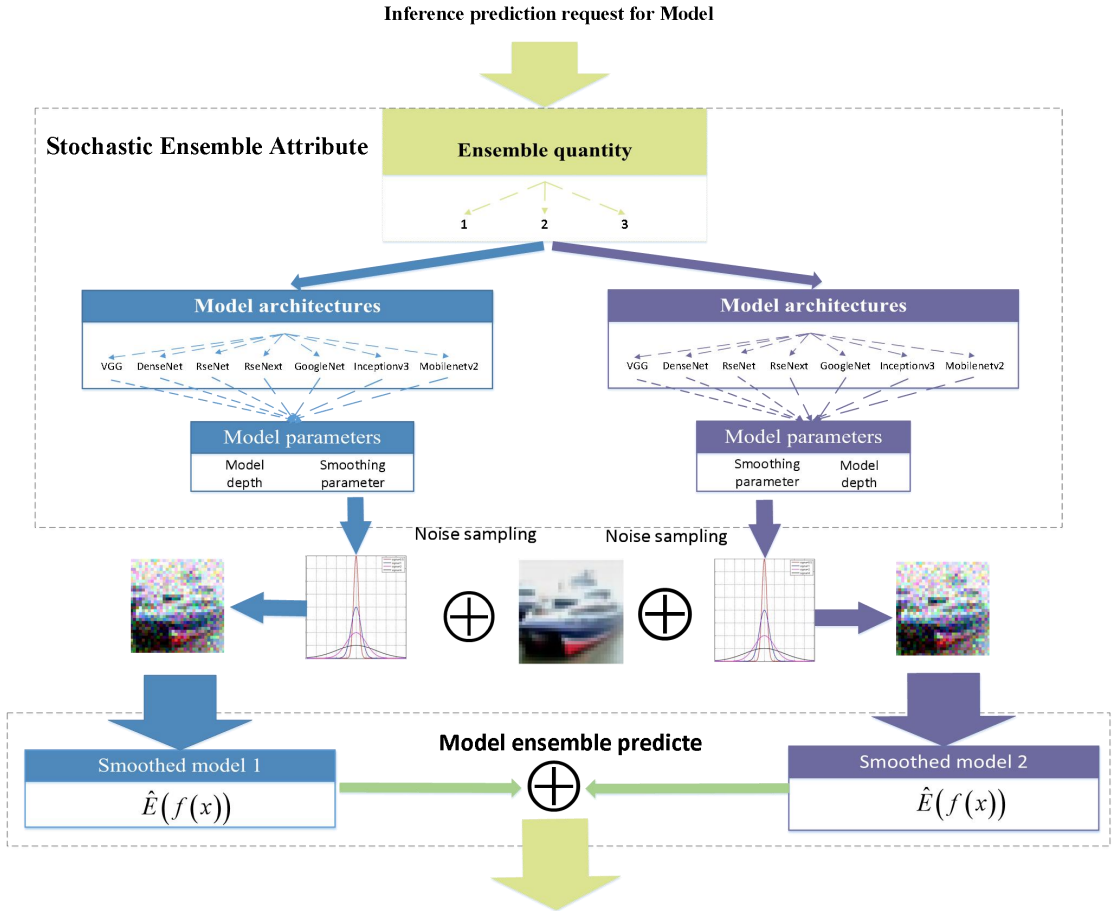

FIGURE 2. A flowchart of the stochastic ensemble smoothing strategy.

The SEM achieves the dynamic defense through the stochastic selection of the ensemble attributes. In summary, the dynamical changes are reflections of the random distribution of the input noise and the uncertified probabilistic gradient information during each ensemble iteration. Essentially, the uncertified ensemble attributes shield the gradient and increase the confusion under white-box and black-box attacks.

## 4. Experiments and Results

Because the uncertified probabilistic gradient of the stochastic ensemble method (SEM) prevents attackers from getting complete model knowledge, there is no clearly defined boundary between the white-box and black-box attacks. Just like risk assessment in cyberspace security, the verification and assessment of adversarial robustness is constrained by the rule: the most capable attacker cannot easily achieve the corresponding attack success rate in the compared model. The attack in this situation might be viewed as a detailed sub-division of the attacker capability, which is used to further describe the adversarial capability in a more global setting. The experiments in our work differ from the random smooth verification of the robustness radius, but includes experiments in the actual adversarial attack environment for evaluation. Instead of the traditional point estimation method, the *ASR-vs-distortion* curves obtained in association with optimal search [48] are used for evaluating the adversarial robustness of the compared defense methods.

4.1 Attack success evaluation metrics based on empirical risk

The *ASR-vs-distortion* curves are created based on precise evaluation of specific adversarial samples during optimal search [48]. Because of the randomness of the ensemble and the input images, the Monte Carlo simulation is used for approximate evaluation in the same fashion as that in random smoothing [29]. Each adversarial sample $x_{adv}$ will be hard predicted $N$ times through the SEM, and the most predicted category will be considered as the output category with an arbitrarily high probability. The baseline classification accuracy of clean sample through this Monte Carlo simulation is 93.4% while the attack success rate for an $N$-time Monte Carlo simulation with the image $x$ is given by:

$$Succ(C, A_{\varepsilon,p}) = \begin{cases} \left(\frac{1}{N}\left(\sum_{n=1}^{N}\left(\sum_{k=1}^{K} g_k(A_{\varepsilon,p}(x))\right)_{one\_hot}\right)\right)_{\max} \neq y & \text{untargeted} \\ \left(\frac{1}{N}\left(\sum_{n=1}^{N}\left(\sum_{k=1}^{K} g_k(A_{\varepsilon,p}(x))\right)_{one\_hot}\right)\right)_{\max} = y_t & \text{targeted} \end{cases} \quad (5)$$

where $A_{\varepsilon,p}$ denotes the adversarial sample with a perturbation budget ε under the $l_p$ norm, and $g_k$ denotes the $k$-th SEM model output. The output of the $N$-time Monte Carlo simulation is the class count statistics obtained by one-hot encoding of the category probability vector. Through the probability conversion function, each predicted value in the vector is converted to the equivalent probability that the SEM puts the adversarial sample into a certain category. Such probabilities can be used in a two-sided hypothesis test that the data conforms to the binomial distribution $n_{succ} \sim Binomial(n_{succ} + n_{nonsucc}, P)$. Monte Carlo simulations of the stochastic ensemble model are conducted to estimate the attack success probability which is redefined as follows:

$$Succ(C, A_{\varepsilon,p}) = \begin{cases} \left(\frac{1}{N}\left(\sum_{n=1}^{N}\left(\sum_{k=1}^{K} g_k(A_{\varepsilon,p}(x))\right)_{one\_hot}\right)\right)_{\max} \neq y \text{ or } \frac{1}{N}\left(\sum_{n=1}^{N}\left(\sum_{k=1}^{K} g_k(A_{\varepsilon,p}(x))\right)_{one\_hot}\right)_{\substack{\max \\ c \neq y}} \geq \alpha & \text{untargeted} \\ \left(\frac{1}{N}\left(\sum_{n=1}^{N}\left(\sum_{k=1}^{K} g_k(A_{\varepsilon,p}(x))\right)_{one\_hot}\right)\right)_{\max} = y_t \text{ or } \frac{1}{N}\left(\sum_{n=1}^{N}\left(\sum_{k=1}^{K} g_k(A_{\varepsilon,p}(x))\right)_{one\_hot}\right)_t \geq \alpha & \text{targeted} \end{cases} \quad (6)$$

In order to control the potential empirical model risk, the abstention threshold α is set to bound the acceptable probability of returning an incorrect prediction [49]. The value of α directly affects the *ASR-vs-distortion* curves making this robust evaluation method more suitable for Monte Carlo simulations. In the following, threshold α is set as 0.3 and the evaluation of the random smoothing model is also based on this metric.

4.2 Attack scenarios

In this section, the attacker's knowledge of the SEM attributes is analyzed in detail. Also,

attack scenarios are designed with different attack capabilities, and the robustness of the proposed method is fully characterized. For the SEM, the attacker can't fully grasp the model gradients, but still can grasp the ensemble attributes and implement transfer-based black-box attacks. The degree of knowledge of the model collection and the ensemble attributes determines the attacker's capabilities. In our work, the attacker's capabilities are defined in a top-down fashion as follows:

1. The attacker *A* has full knowledge of the model collection, and has the ability to obtain the ensemble attributes in real time but cannot forecast these attributes for the next iteration. (The attacker can detect the gradient changes in each iteration of a white-box attack.)
2. The attacker *B* also has complete knowledge of the model collection but obtains the attributes periodically. (The gradient of the model does not change in each iteration of a white-box attack.)
3. The attacker *C* has a part of the model collection (half of all model library in experiment) and tries to use the SEM as regularization constraints to improve the transferability of the adversarial samples.
4. The attacker *D* has a part of the model collection and tries to use the traditional ensemble method to generate adversarial samples.
5. The attacker *E* has no knowledge whatsoever of the model collection and gradients, and only obtains the model probability vector to perform a source-based black-box attack.

In this paper, the traditional method is used as the baseline model for contrast. Based on the attack type and the defense method, eight baseline models can be identified:

1. The ensemble smoothed model under a white-box attack (the contrast model *F*)
2. The smoothed model under a white-box attack (the contrast model *G*)
3. The ensemble model under a white-box attack (the contrast model *H*)
4. The single model under a white-box attack (the contrast model *I*)
5. The ensemble smoothed model under a black-box attack (the contrast model *J*)
6. The smoothed model under a black-box attack (the contrast model *K*)
7. The ensemble model under a black-box attack (the contrast model *L*)
8. The single model under a black-box attack (the contrast model *M*)

4.3 Robustness analysis based on the attack scenario

A more detailed evaluation of adversarial robustness can be achieved through considering the combinations of different attack capabilities, attack methods, attack targets, and disturbance constraints. So, we further carry out transfer-based attacks on ensemble models using three standard methods of attack, namely, the basic iterative method (BIM), the momentum iterative method (MIM) and projected gradient descent method (PGD) with each of the attackers *A*, *B*, *C*, *D* and baseline model *F, G, H, I*. The BIM is proposed for the white-box attacks while the MIM is proposed for the transfer-based black-box attacks. The two methods differ in the transferability improvement, and the associated results show the influence of transferability for adversarial robustness. We also use the natural evolutionary strategy (NES) attack and the simple perturbation stochastic approximation (SPSA) attack in conjunction with the attackers *E, J, K, L* and *M*. The NES and SPSA attacks commonly estimate gradient information through different loss distribution hypotheses. Experimental comparisons show the influence of the SEM probabilistic outputs on gradient estimation. For all *ASR-vs-distortion* curves, the search step was set to 10 while the binary search step was set to 20. For the white-box attacks, the number of attack iterations of both the BIM and MIM was set to 20, while the maximum queries count was set to 5000 for the source-based black-box attacks

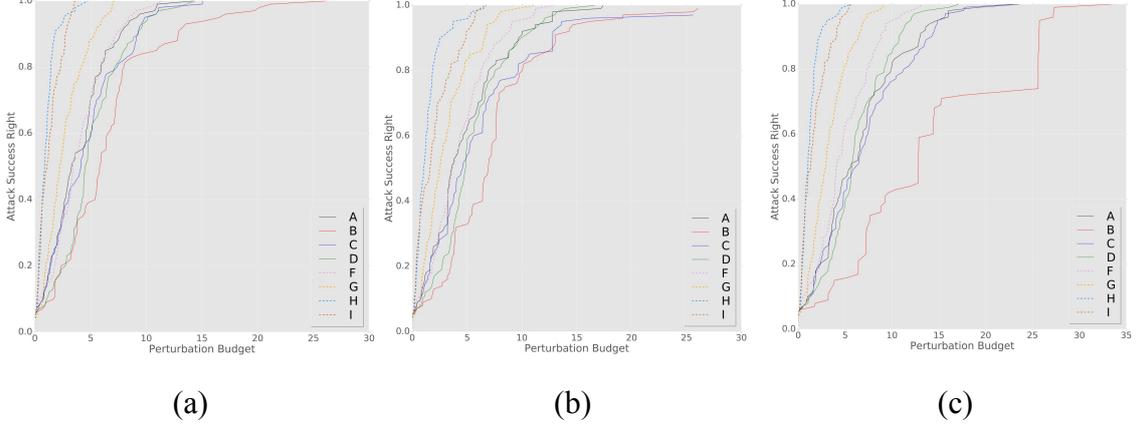

(a) (b) (c)

FIGURE 3. The *ASR-vs-distortion* curves for untargeted transfer-based white-box attacks: (a) BIM, (b) MIM, (c) PGD

Figure 3 shows the *ASR-vs-distortion* curves for untargeted transfer-based white-box attacks with the SEM on the CIFAR-10 dataset[50]. *A, B, C* and *D* shows the different attack scenarios while the baseline contrast *F, G, H* and *I* are shown in dotted curve. By comparing the baseline models, we can see that the ensemble model is extremely vulnerable under white-box attacks, which is worse than single model. Smoothing method improves the robustness of a single model, and the ensemble smoothed model further improves the robust performance and solves the vulnerability of ensemble under white-box attacks. For all the attack method, the attackers *B* has worst attack performance, showing that the ability to protect model from grasping gradient information frequently of each iteration is significant for SEM robustness. Nevertheless, attacker *D*, which has the part knowledge of candidate model library but knowledge of the ensemble attributes in each iteration, achieves the transfer attack though ensemble and achieve the similar robustness performance (even exceed in PGD) with attacker *A* who has the most knowledge of SEM. Compare the performance of *C* with *D*, the SEM does not improve the attack effect as a regularization method. This reveals the importance of protecting the model library for the SEM robustness. For the smoothed ensemble model which has the best robustness performance between baseline models, the best attack performance cannot easily exceed the attack success rate associated with it. When the attacker has higher transferability (for the MIM and PGD), the advantages of the transferability is only for attacker D but no longer attained through SEM.

Figure 4 shows the *ASR-vs-distortion* curves for targeted transfer-based white-box attacks. Comparing different attack methods, the improved transferability of the PGD method does not significantly improve the attack performance of SEM, but the defense effect is significantly improved on the momentum-based attack, which shows that the randomness of the gradient has a certain effect on the confusion of the gradient direction. *A* and *C* have tiny difference in robustness performance, showing that the little different grasp of model library cannot influence the robustness of SEM under targeted attack. But different from conclusion under untargeted attack, the robustness performance under *A* and *C* do not always have better robustness performance than that under ensemble smoothed or single smoothed model demonstrating the lack of heterogeneity of the model in the gradient direction. We believe that a model smoothed under the same data set will mildly change the decision boundary due to factors such as ensemble architecture, resulting in a difference in the recent non-category gradient direction, so as to effectively defend through SEM. But because the training data is the same, the gradient direction of a specific category will not be different enough through different model architecture, making SEM not always as robust as the ensemble smoothed model. However, through the detail result in the second line of Figure 4, the proposed method always achieve a better robustness with small perturbation. Comparing the attackers *A*, *B*, *C* and *D* we realize that the worst attack performance is that of *B* (This situation for attacker *D* might be reversed with the PGD attack method). Further combined with the results of untargeted attacks, we think the decrease of ensemble change frequency is important for the SEM when the model library and ensemble attribute can be gained by attacker. In the case of small disturbance, the SEM achieves good robustness.

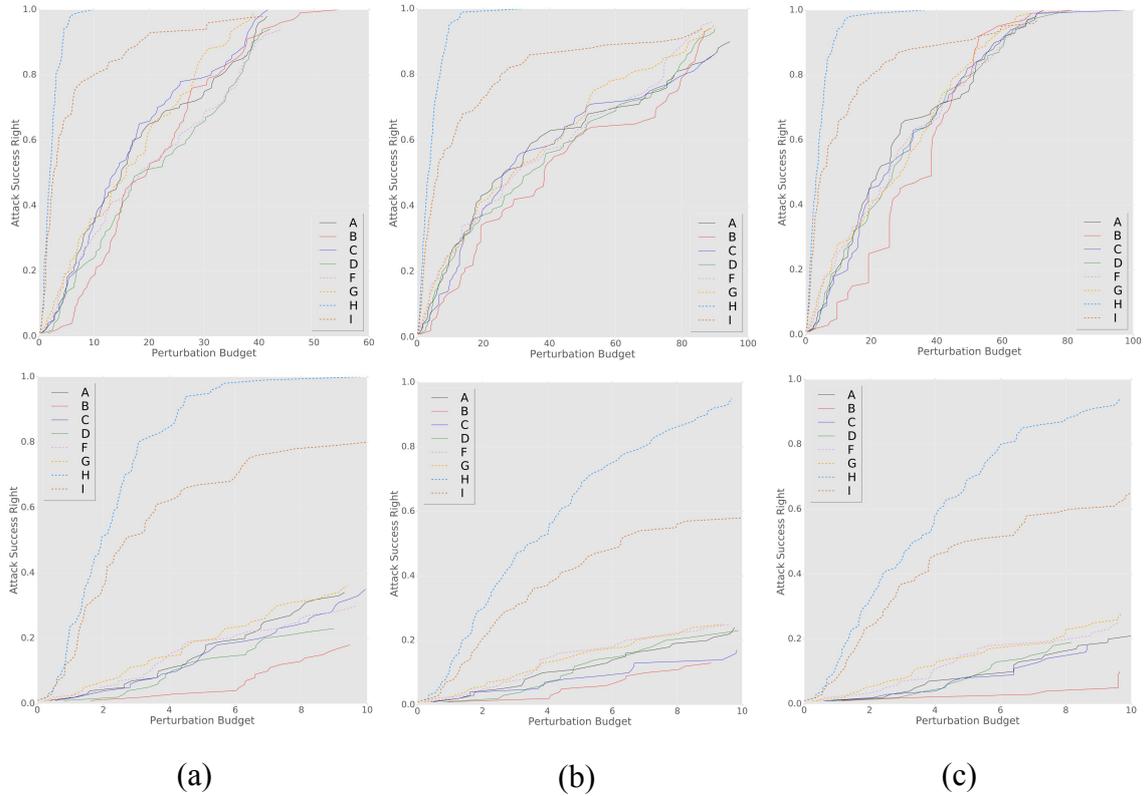

(a)          (b)          (c)

FIGURE 4. The *ASR-vs-distortion* curves for targeted transfer-based white-box attacks: (a) BIM,

(b) MIM, (c) PGD; Second line shows the detail result with small perturbation

Through overall analyzing of robust performance under the transfer-based white box attack, SEM comprehensively enhances the robustness under white-box untargeted attacks. Under white-box targeted attacks, certain additional production method need to be taken to effectively control the attackers' ability so as to achieve good robustness. In general, the SEM achieves good robustness in the case of small disturbance. When the attacker has a highest ability to know the model library and ensemble attribute, reducing the change frequency of SEM or preventing attacker from obtaining ensemble attribute is more helpful to confuse the attacker. If the attacker does not know the ensemble attribute, the better robustness of the SEM can be achieve by protecting the model library and further expanding the heterogeneity of the model library more than architecture.

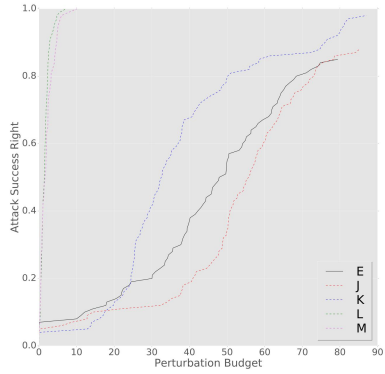
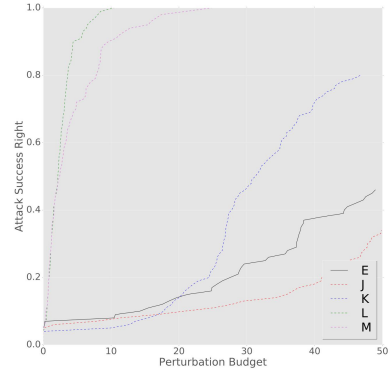

(a)                        (b)

FIGURE 5. The *ASR-vs-distortion* curves for untargeted source-based black-box attacks: (a) NES, (b) SPSA

Figure 5 shows the results of an untargeted source-based black-box attack. Obviously, the ensemble model robustness is slightly worse than that of the single model both for NES and SPSA attacks showing the vulnerability of ensemble model under black box. Both the SPSA and NES methods assume that the gradient direction of the adversarial sample conforms to a specific probability distribution. The assumption of gradient direction is accumulated by randomly sampling under a probability distribution, whose step is controlled by the loss value. Under such an expectation hypothesis based on the probability of the gradient distribution, the evaluation of the SEM is essentially to evaluate the degree of overlap between the gradient direction of the model and the assumed distribution direction under the probability. In the experiment, SEM does not show a better untargeted black box defense effect than smoothed ensemble, indicating that SEM based on different smoothing parameters are more likely to be affected by noise expectations with high variance $\sigma$ (set as 1 in experiment). We think that this property is due to the high ensemble probability of a smoothed model with low variance $\sigma$ or non-smoothed model, so the defense effect of SEM is not as good as the ensemble smoothed model in terms of probability.

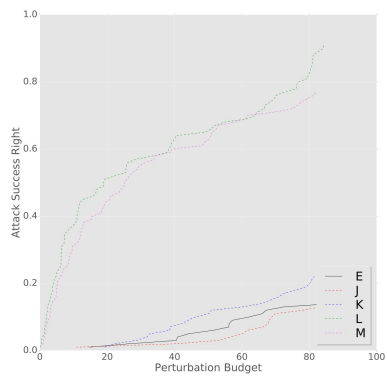
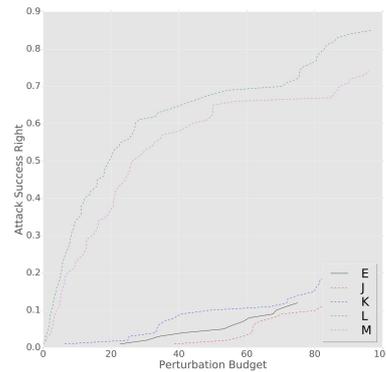

(a)                        (b)

FIGURE 6. The *ASR-vs-distortion* curves for targeted source-based black-box attacks: (a) NES, (b) SPSA

In comparison, the results for the targeted source-based black-box attacks are shown in Figure 6. In terms of overall accuracy, targeted attacks have all declined but the same conclusion about robustness between different models can be obtained. Through experiments on the model under black-box attacks, the sensitivity of the model to the specific distribution noise is analyzed. Because of the smoothing of the model, it has better defense performance against adversarial samples based on specific noise distribution assumptions. However, the model under different smoothing parameters is sensitive to the noise with different parameters, which also limits the defense performance. Such noise assumptions are independent of the true gradient information of the model, and mainly depend on the change of model output and the number of queries. Further improvement of defense effect may need to rely on the optimization of the smooth parameter selection of the ensemble strategy.

4.4 Robustness analysis based on the stochastic ensemble strategy

For the SEM stochastic ensemble strategy, we investigate here the impact of the model quantity and the model heterogeneity on the SEM robustness. For the ensemble model quantity, the optional values of 1, 2, and 3 are changed to become 6, 7, and 8, respectively for the compared model. For the heterogeneous ensemble attributes, a stochastic ensemble of a single-architecture neural network based on different smoothing parameters is used as a comparison model. Because of the good baseline and smoothed prediction accuracies of the WRN [40] shown in Table 1, this network is used herein as the single-architecture neural network. To construct a space of the stochastic ensemble model collection that is as big as that of the SEM, and also to create model gradient variations, the WRN [40] resorted to smoothing by Gaussian noise (where we set $\delta$ as 0.12, 0.15, 0.25, 0.5, 0.75, 1.0 and 1.25) using different training methods (stability training [30], semi-supervised learning [51] and pre-training [47]). The adversarial robustness values of the different ensemble strategies were evaluated for the white-box attacker $A$ in order to show the gradient shielding effects of these strategies.

Figure 7 and Figure 8 show the robustness evaluation results under different strategies. The impact of the ensemble model quantity on the robustness is mainly visible for all attacks aim. As we mentioned in the method part, a larger model quantity of the ensemble will reduce the gradient differences and also improve the transferability of adversarial samples in each ensemble iteration. As Figure 7 shows in blue, the architectural heterogeneity of the model collection has a bigger impact on the SEM adversarial robustness. When there are no architectural differences between the ensemble models, even in the stochastic smoothed case, the stochastic ensemble becomes a method to increase the SEM vulnerability to the adversarial samples under both targeted and untargeted attacks. As shown in Figure 8, the stochastic ensemble without heterogeneous model will even more vulnerable than an ensemble model. Through viewing the SEM ensemble strategy as a kind of dropout operation [52], if the ensemble model quantity is large and there is no enough architectural differences in the model collection, the SEM method reduces to a kind of a regularization method conversely improving the ability of the adversarial samples especial under targeted attack.

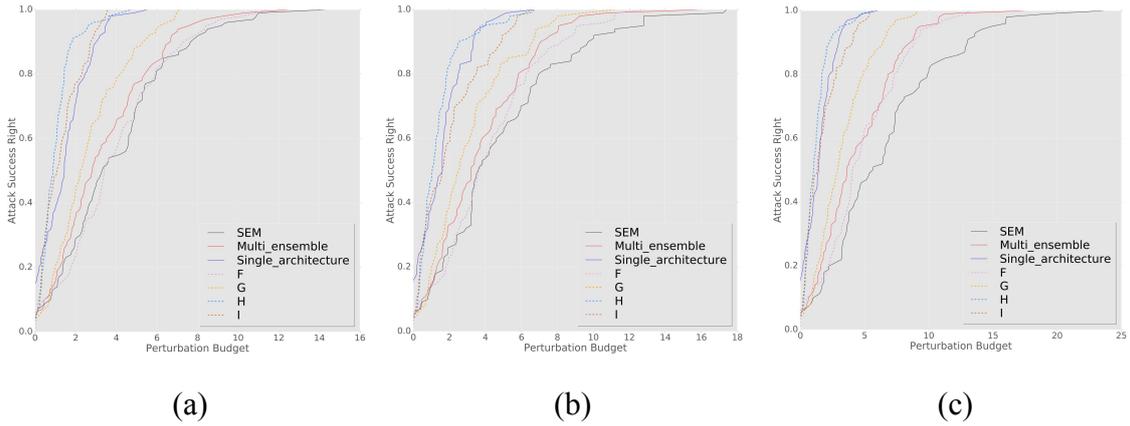

FIGURE 7. The *ASR-vs-distortion* curves for white-box untargeted attacks under different attack methods and ensemble strategies: (a) BIM, (b) MIM, (c) PGD

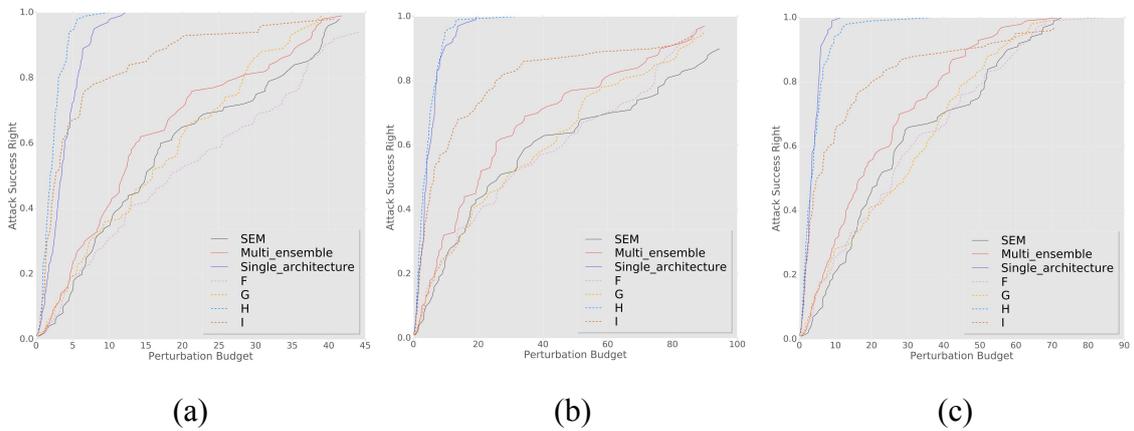

FIGURE 8. The *ASR-vs-distortion* curves for white-box targeted attacks under different attack methods and ensemble strategies: (a) BIM, (b) MIM, (c) PGD

## 5. Conclusions

This paper utilizes the ensemble smoothed model to create dynamic defense methods for generalized robustness control of deep neural networks. The ensemble attributes are regard as the changeable factor and dynamically adjusted during the inference prediction phase, so that the attack method cannot effectively iterate in a specific gradient direction under white-box attack. Through optimal search for the perturbation values under different attack capabilities, attack methods and attack targets according to the degree of the real-time capability of an attacker to obtain knowledge of the model collection and gradients, a comprehensive evaluation demonstrates that: when the image distortion is small, even the attacker with the highest attack capability cannot easily exceed the attack success rate associated with the ensemble smoothed model, especially under untargeted attacks. The heterogeneity and confidentiality of the model collection play a crucial role in the robustness of the proposed method. The proposed method has the characteristics of diversity, randomness, and dynamics to achieve the probabilistic attribute dynamic defense for adversarial robustness. In order to further improve the robustness under the white-box attack, the frequency of ensemble change can be adaptively controlled by attack iteration. For the robustness under black-box attack, further improvement of defense effect may need to rely on the optimization of the smooth parameter selection of the ensemble strategy and model library with more weak classifiers smoothed .under higher parameter $\sigma$ .